\documentclass[twocolumn,letterpaper,10pt]{article}  %%    [COMMENT SS 2C]
\pdfoutput=1
\usepackage[dvips,pdftex]{graphicx}
\usepackage[]{caption}
\usepackage{morefloats}
\usepackage{rotating}
\usepackage{subcaption}
\usepackage{array}
\usepackage{tabularx}
\usepackage{epstopdf}
\usepackage{enumitem}
\usepackage{seqsplit}
\usepackage{amsmath}
\usepackage{latexsym}
\usepackage{amssymb}
\usepackage[numbers,super,sort&compress]{natbib}
\usepackage{color}
\usepackage[rgb,dvipsnames]{xcolor}
\definecolor{grey}{gray}{0.7}
\usepackage{hyperref}
\usepackage[compact]{titlesec}
\usepackage{fancyhdr}
\usepackage[paperwidth=8.5in, paperheight=11.0in, textwidth=7.00in, textheight=9.00in,
bindingoffset=0pt, includehead=false, includefoot=false, ignoreheadfoot=true, layoutsize={8.5in,11.0in}, layoutoffset={0pt,0pt}, hcentering=true]{geometry}
\usepackage{setspace}
\usepackage{indentfirst}
\usepackage{lastpage}     %%  Allow user to know total # of pages (e.g. page 5 of 9)
\hypersetup{colorlinks=true, urlcolor=blue, citecolor=black, linkcolor=black}
\setlist{leftmargin=3.5mm}
\newcommand{\papertitle}{Relativistic electrons produced by foreshock disturbances}
\newcommand{\headertext}{WILSON ET AL.}
\newcommand{\mpageofn}{\thepage\ of \pageref{LastPage}}
\newcommand{\evenhdrnote}{Phys. Rev. Lett.}
\newcommand{\oddhdrnote}{\headertext \ \mpageofn}
\newcommand{\affilone}{NASA Goddard Space Flight Center, Heliospheric Physics Laboratory, Greenbelt, MD 20771, USA}
\newcommand{\affiltwo}{The Aerospace Corporation, El Segundo, CA 90245, USA}
\newcommand{\affilthr}{Department of Radio Science, Aalto University, Aalto, Finland}
\newcommand{\affilfou}{Department of Astrophysical Sciences, Princeton University, Princeton, NJ, USA}
\newcommand{\affilfif}{Department of Earth and Space Sciences, University of California, Los Angeles, Los Angeles, CA, USA}
\DeclareCaptionLabelSeparator{vline}{$\boldsymbol{\lvert}$ }
\newcommand{\footnoteremember}[2]{\footnote{#2}\newcounter{#1}\setcounter{#1}{\value{footnote}}}
\newcommand{\footnoterecall}[1]{\footnotemark[\arabic{#1}]}
\begin{document}
%%\linenumbers                 %%  turn on line numbering
%%  \runninglinenumbers  %%  force running line numbers  [COMMENT SS 2C]
%%\runningpagewiselinenumbers  %%  force running numbers in pagewise mode  [COMMENT SS 2C]
%%  style of bibliography
%%  \bibliographystyle{naturemag}
%%  Vertical spacing before/after each
\titlespacing{\enumerate}{0pt}{0pt plus 1pt minus 1pt}{0pt plus 1pt minus 1pt}
\titlespacing{\figure}{0pt}{0pt plus 1pt minus 1pt}{0pt plus 1pt minus 1pt}
\titlespacing{\table}{0pt}{0pt plus 1pt minus 1pt}{0pt plus 1pt minus 1pt}
\titlespacing{\wrapfigure}{0pt}{0pt plus 1pt minus 1pt}{0pt plus 1pt minus 1pt}
%% => Shut off section counters
\setcounter{secnumdepth}{-1}
%%
%%  Format captions
%%
\captionsetup[figure]{labelfont=bf,skip=1pt,font=footnotesize,position=below,justification=RaggedRight,labelsep=vline}
%%
%%  Page Format and Numbering
%%
\pagenumbering{arabic}  %%  start page numbering here
\setcounter{page}{1}
%%
%%  Title Page
%%
\title{\bf \papertitle}
\author{Lynn B. Wilson III\footnoteremember{1}{\affilone}, David G. Sibeck\footnoterecall{1}, Drew L. Turner\footnoteremember{2}{\affiltwo}, Adnane Osmane\footnoteremember{3}{\affilthr}, Damiano Caprioli\footnoteremember{4}{\affilfou} \\ \& Vassilis Angelopoulos\footnoteremember{5}{\affilfif}}
\date{\today}
\maketitle
%%----------------------------------------------------------------------------------------
%%  Intro Paragraph [length ≤ 300 words]
%%    1)  1-2 sentences providing basic intro [comprehensible to any discipline]
%%    2)  2-3 sentences of more detailed background [comprehensible to your discipline]
%%    3)  1 sentence stating the problem addressed
%%    4)  1 sentence summarising the main result [e.g., "Here we show..."]
%%    5)  2-3 sentences explaining the main result
%%        A)  directly compare to previous ideas; or
%%        B)  how main result adds to previous knowledge
%%    6)  1-2 sentences putting results into a more general context
%%    7)  2-3 sentences providing a broader perspective
%%
%%----------------------------------------------------------------------------------------
\noindent  \textbf{Foreshock disturbances -- large-scale ($\sim$1000 km to $>$30,000 km), \textcolor{Red}{solitary} ($\sim$5--10 per day\textcolor{Red}{, transient (lasting $\sim$}10s of seconds to several minutes\textcolor{Red}{)}) structures\cite{omidi07a, turner13a} -- generated by suprathermal ($>$100 eV to 100s of keV) ions\cite{burgess12a, wilsoniii16a} arise upstream of Earth's bow shock formed by the solar wind colliding with the Earth's magnetosphere.  They have recently been found to accelerate ions to energies of several keV\cite{kis13a, wilsoniii13b}\textcolor{Red}{.  One type was found to have a distinct suprathermal electron population with energies $>$70 keV, which was attributed to a magnetospheric origin\cite{paschmann88a}.}  Although electrons in Saturn's high Mach number (M $>$ 40) bow shock can be accelerated to relativistic energies (nearly 1000 keV)\cite{masters13a}, it has hitherto been thought impossible to accelerate electrons at the much weaker (M $<$ 20) Earth's bow shock beyond a few 10s of keV\cite{wu84b}.  Here we report observations of electrons energized by foreshock disturbances \textcolor{Red}{from 10s of eV} up to at least $\sim$300 keV.  \textcolor{Red}{We observe a single isotropic power-law from 100s of eV to 100s of keV, unlike previous studies\cite{paschmann88a}.  All previous observations of energetic foreshock electrons have been} attributed to escaping magnetospheric particles\cite{paschmann88a, krimigis78a, sarris87a} or solar events\cite{ergun98e}.  \textcolor{Red}{We observe no solar activity and the single isotropic power-law cannot be explained by any magnetospheric source.}  Further, current theories of ion acceleration in foreshock disturbances cannot account for electrons accelerated to the observed relativistic energies\cite{anagnostopoulos98a, caprioli14a, malkov01a, park13a, park15a, treumann09a}.  These electrons are clearly coming from the disturbances, leaving us with no explanation \textcolor{Red}{for the acceleration mechanism}.}

%%----------------------------------------------------------------------------------------
%%  Introduction:  Data Sets
%%----------------------------------------------------------------------------------------
\indent  We examine in detail particle velocity distribution functions measured by the low energy electron/total ion electrostatic analyzers\cite{mcfadden08a} and high energy electron/total ion solid state telescopes\cite{ni11b} on the THEMIS spacecraft\cite{angelopoulos08a} near foreshock disturbances.  Quasi-static vector magnetic fields ($\mathbf{B}{\scriptstyle_{o}}$) were measured by the fluxgate magnetometer\cite{auster08a}.

%%  Introduction:  Short study summary
\indent  The disturbances occur in the ion foreshock\cite{burgess12a, wilsoniii16a} region upstream of the quasi-parallel bow shock, where the shock normal angle ($\theta{\scriptstyle_{Bn}}$) between the upstream quasi-static magnetic field ($\mathbf{B}{\scriptstyle_{o}}$) and the shock normal vector satisfies $\theta{\scriptstyle_{Bn}}$ $<$ 45$^{\circ}$.  We focused on observations near short large-amplitude magnetic structures\cite{wilsoniii16a}, hot flow anomalies\cite{omidi07a}, and foreshock bubbles\cite{turner13a}.  During the four orbital passes examined in detail, we identified 30 foreshock disturbances, 10 of which had clear energetic electron enhancements \textcolor{Red}{(five at short large-amplitude magnetic structures, two at hot flow anomalies, and three at foreshock bubbles)}.  See Extended Data Fig. \ref{fig:THEMISOrbits} for spacecraft orbits and Methods for further details of foreshock disturbances.

%%----------------------------------------------------------------------------------------
%%  Observations:  Introduce figures
%%----------------------------------------------------------------------------------------
%%  Figure 1
\indent  We observe energetic ($\geq$30 keV) electron enhancements as short duration ($\sim$10s of seconds to few minutes) enhancements in the electron fluxes above background by factors of $\sim$10--200 (Fig. \ref{fig:exampleTIFPs}).  They are localized to \textcolor{Red}{the} large fluctuations in $\mathbf{B}{\scriptstyle_{o}}$ (Figs \ref{fig:exampleTIFPs}\textbf{a}--\ref{fig:exampleTIFPs}\textbf{c}) within the foreshock disturbances.  The electron flux-time profiles for energies from $\sim$0.25 to $>$200 keV show no energy dispersion, i.e., fluxes increase at all energies simultaneously (Figs \ref{fig:exampleTIFPs}\textbf{g}--\ref{fig:exampleTIFPs}\textbf{i} and \ref{fig:exampleTIFPs}\textbf{m}--\ref{fig:exampleTIFPs}\textbf{o}).  The low energy electron (Figs \ref{fig:exampleTIFPs}\textbf{g}--\ref{fig:exampleTIFPs}\textbf{i}) and ion (Figs \ref{fig:exampleTIFPs}\textbf{j}--\ref{fig:exampleTIFPs}\textbf{l}) data look qualitatively similar for disturbances with and without (Extended Data Fig. \ref{fig:exampleTIFPswithoutee}) energetic electron enhancements.  Note that nearly all disturbances show enhanced energetic ion fluxes often to $>$300 keV (Figs \ref{fig:exampleTIFPs}\textbf{p}--\ref{fig:exampleTIFPs}\textbf{r}).  The ion enhancements will be examined in future work.

%%  Figure 2
\indent  We then constructed pitch-angle -- \textcolor{Red}{the} angle between electron momentum and some reference vector, often $\mathbf{B}{\scriptstyle_{o}}$ -- distributions from the particle data to compare distributions during energetic electron enhancements (Figs \ref{fig:examplePADs}\textbf{a}--\ref{fig:examplePADs}\textbf{c}) to those outside the enhancement periods (Figs \ref{fig:examplePADs}\textbf{d}--\ref{fig:examplePADs}\textbf{f}).  The electron distributions are observed to be almost isotropic from $\sim$0.05--300 keV, with anisotropies (e.g., ratio of intensity or phase space density along $\mathbf{B}{\scriptstyle_{o}}$ to that perpendicular to $\mathbf{B}{\scriptstyle_{o}}$) rarely exceeding factors of 2.  Note that data above $\sim$140 keV \textcolor{Red}{are} dominated by noise for the 2008-09-08 hot flow anomaly (Fig. \ref{fig:examplePADs}\textbf{b}) and 2008-07-14 foreshock bubble (Fig. \ref{fig:examplePADs}\textbf{c}) example enhancements.  The data also show a power-law form with $f\left(E\right)$ $\propto$ $E^{-4}$ from as low as $\sim$0.25 keV up to highest energies observed during each enhancement (Figs \ref{fig:examplePADs}\textbf{a}--\ref{fig:examplePADs}\textbf{c}).  The distributions observed outside the enhancements (Figs \ref{fig:examplePADs}\textbf{d}--\ref{fig:examplePADs}\textbf{f}) show far more variability, only noise $>$12 keV, and in some cases significant anisotropies (Fig. \ref{fig:examplePADs}\textbf{f}).

%%  Aliasing
\indent  Due to the high variability in $\mathbf{B}{\scriptstyle_{o}}$ during the accumulation time of a single particle distribution we also constructed pitch-angle distributions by sorting the pitch angles with sub-spin period time resolution, using algorithms similar to previous work\cite{wilsoniii12c}, to remove artifacts of aliasing.  We found that the isotropy was not a consequence of aliasing and is thus a real feature (see Extended Data Fig. \ref{fig:exampleCompareLTHTBo}).

%%  Solar source
\indent  To verify that the enhancements were not of solar origin, we examined the \emph{Wind} and \textcolor{Red}{twin} STEREO spacecraft radio data for solar radio bursts, which were not observed during the energetic electron enhancements observed by THEMIS.  We also examined the \emph{Wind} particle data for energy-dispersed profiles characteristic of solar energetic electrons\cite{ergun98e}, which were also not observed.  Finally, there were no interplanetary shocks, which can produce relativistic electrons\cite{desai08a, sarris85a}, observed by \emph{Wind} during any of the electron enhancements.  Further details excluding a solar source can be found in Methods and see Extended Data Fig. \ref{fig:WindandSTEREOWAVESRadio}.

%%  Bow shock source
\indent  Next we ruled out the Earth's bow shock as a source by examining the following concepts.  First, the two most viable shock acceleration mechanisms for electrons have the highest efficiencies in the quasi-perpendicular ($\theta{\scriptstyle_{Bn}}$ $>$ 45$^{\circ}$) region of the shock\cite{wu84b, park13a}.  Second, due to a higher mobility along versus across $\mathbf{B}{\scriptstyle_{o}}$, any energetic electron distribution observed several Earth radii (i.e., tens to hundreds of Larmor radii) away from the source would be highly anisotropic along $\mathbf{B}{\scriptstyle_{o}}$, as previously reported\cite{anderson79a}, inconsistent with our observations (e.g., Figs \ref{fig:examplePADs}\textbf{a}--\ref{fig:examplePADs}\textbf{c}).  Third, the maximum observed energies are at most several tens of keV energies\cite{desai08a, gosling89b, sarris85a} seen as magnetic field-aligned beams at the electron foreshock edge, whereas we observe isotropic distributions up to at least $\sim$300 keV within the ion foreshock (Figs \ref{fig:exampleTIFPs} and \ref{fig:examplePADs}).  Finally, both of these mechanisms should produce some energy dispersion for upstream observations, which we do not observe (Fig. \ref{fig:exampleTIFPs}).  For a detailed discussion of these mechanisms see Methods and Extended Data Fig. \ref{fig:THEMISFBPADExample}.

%%  pile up source
\indent  Another possibility is that the enhancements arise as a ``pile up'' of particles at the magnetic gradients, which act like mirrors.  However, we do not observe energy-dispersed profiles within the magnetic compressions (i.e., due to energy-dependent Larmor radii effects and diffusion) or enhanced intensities outside the foreshock disturbances or pitch-angle distributions peaking perpendicular to $\mathbf{B}{\scriptstyle_{o}}$.  Further, the Larmor radius of a $\sim$90 keV electron, in the range of magnetic field magnitudes observed, span $\sim$26--1080 km.  These values are comparable to and larger than the gradient scale length of most collisionless shock waves\cite{wilsoniii16a} and comparable to the scale size of some foreshock disturbances, which precludes adiabatic reflection between the two merging shocks\cite{treumann09a}.  Thus, we can rule out the ``pile up'' scenario as a source.

%%  Magnetospheric source
\indent  \textcolor{Red}{Previous} reports of energetic electrons in the ion foreshock attributed the enhancements to a magnetospheric source\cite{anagnostopoulos98a, krimigis78a, sarris87a}.  \textcolor{Red}{One study\cite{paschmann88a} found energetic electrons ($\sim$70--200 keV) within hot flow anomalies that were nearly isotropic, but the authors explicitly stated that the energetic electrons were a separate population from those at lower energies and that they were of magnetospheric origin.  There are multiple differences between these results and our observations, including:  (1) while hot flow anomalies can be magnetically connected to the magnetosheath, the other two foreshock disturbances in our study are not; (2) we observe a single power-law from 100s of eV to $\geq$140 keV in many enhancements, indicative of a common acceleration mechanism; and (3) the combination of a single power-law and an isotropic distribution over such a large range of energies cannot be explain by a magnetospheric source.}

\indent  \textcolor{Red}{Finally, the} enhancements were often observed with geomagnetic activity\cite{anagnostopoulos98a, krimigis78a, kronberg11a, sarris87a} and exhibited large anisotropies (ratios of sunward-to-anti-sunward fluxes\cite{anagnostopoulos86a} up to 5:1 and field-aligned-to-perpendicular fluxes\cite{sarris87a} up to 6:1).  \textcolor{Red}{Our observations are not consistent with any of these studies for the following reasons.}  First, since our \textcolor{Red}{electron distributions} are nearly isotropic, they do not exhibit either highly field-aligned (Fig. \ref{fig:examplePADs}) nor anti-earthward intensity anisotropies (see Extended Data Fig. \ref{fig:THEMISFBPADExample}).  Second, we do not observe enhanced geomagnetic activity before or during any of the energetic electron enhancements (see Extended Data Fig. \ref{fig:THEMISAEIndices}).  Finally, the enhancements are always isolated within the foreshock disturbances, thus not consistent with escaping magnetospheric particles.  For further details, see Methods.

%%----------------------------------------------------------------------------------------
%%  Conclusions:
%%----------------------------------------------------------------------------------------
%%  Summary
\indent  No previous work ever considered that foreshock disturbances could locally produce electrons to these energies.  It is not clear why only some disturbances produce enhancements nor can current theory explain the electron energization, but this is beyond the scope of this study.

\indent  The electron energization is not occurring locally at the main shock, but remotely.  The most outstanding unanswered question in shock acceleration theory is the so called ``injection problem'' (i.e., how to get thermal particles up to suprathermal energies before they are convected downstream), where previous work has only considered local energization at the shock.  Therefore, these observations provide a new avenue through which electrons can be non-locally pre-energized to high enough energies to undergo further acceleration when interacting with astrophysical shocks.  Given the ubiquity of foreshocks upstream of collisionless astrophysical shocks, we expect foreshock disturbances to be ubiquitous as well, which could fundamentally change our understanding of collisionless shocks.

%%----------------------------------------------------------------------------------------
%%  Online Content
%%----------------------------------------------------------------------------------------
{\small
{\noindent \textbf{Online Content}} Methods, along with any additional Extended Data display items and Source Data, are available in the online version of the paper; references unique to these sections appear only in the online paper.
}

%%----------------------------------------------------------------------------------------
%%  Bibliography
%%----------------------------------------------------------------------------------------
\vspace{-5pt}
\renewcommand{\bibsep}{0pt}	% Tighten spacing between references
%%  {\footnotesize
%%  \bibliography{/Users/lbwilson/Desktop/Lynn_B_Wilson_III/LaTeX/Bibliographies/my_bib_maker}
%%  }

%%----------------------------------------------------------------------------------------
%%  Acknowledgements
%%----------------------------------------------------------------------------------------
{\small
{\noindent \textbf{Acknowledgements}} A.F.- Vi{\~n}as, D. Bryant, V. Krasnoselskikh, M. Desai, B. Randol, J. Giacalone, F. Guo, A.W. Breneman, and E.R. Christian for useful discussions of the fundamental physics involved in our study.  The authors thank the CARISMA team (operated by the University of Alberta and funded by the Canadian Space Agency), the \emph{Wind} and STEREO teams, and NASA SPDF for data.
}

%%----------------------------------------------------------------------------------------
%%  Author Contributions
%%----------------------------------------------------------------------------------------
%%  authors are required to include a statement of responsibility that specifies the contribution of each co-author.
{\small
{\noindent \textbf{Author Contributions}}  L.B.W.III performed the bulk of the data retrieval and analysis in addition to manuscript and figure preparation.  D.L.T. and D.G.S. identified the foreshock disturbances in addition to providing useful feedback during manuscript preparation.  V.A. validated the THEMIS observations and contributed to data interpretation.  A.O. and D.C. interpreted the results in the context of current theoretical work and simulation results.  All authors contributed to the creation of the manuscript.
}

%%----------------------------------------------------------------------------------------
%%  Author Information
%%----------------------------------------------------------------------------------------
%%  includes a data deposition statement (where relevant), a reprints and permissions statement, a statement declaring competing financial interests (if applicable) and a correspondence line with an e-mail address.
{\small
{\noindent \textbf{Author Information}}  The authors declare no competing financial interests.  Correspondence and requests for materials should be addressed to L.B.W. (lynn.b.wilsoniii@gmail.com).
}

\clearpage
%%----------------------------------------------------------------------------------------
%%  Section:  Methods
%%----------------------------------------------------------------------------------------
\section{Methods}  \label{sec:Methods}

%%  Subsection:  Foreshock Disturbance Properties
{\noindent \textbf{Foreshock disturbance properties.}}  Since previous simulation\cite{riquelme11a} and observational\cite{oka06a, wilsoniii12c} studies found that magnetosonic-whistler waves\cite{wilsoniii16a} -- right-hand polarized electromagnetic plasma waves, with density fluctuating in phase with $\lvert \mathbf{B}{\scriptstyle_{o}} \rvert$ -- can accelerate particles to $>$1 keV, we limited the types of foreshock disturbances to those having a magnetosonic-whistler nature.  We further limited our search to disturbances occurring multiple times per day, those with nonlinear properties, and those capable of forming shock waves, all properties associated with energizing particles.  Therefore, the three foreshock disturbances we examined, short large-amplitude magnetic structures\cite{lucek08a, mann94a, scholer03b, schwartz92a, wilsoniii13b, wilsoniii16a}, hot flow anomalies\cite{eastwood08a, omidi07a, omidi13a, omidi14c, schwartz85a, zhang10a, zhang13a}, and foreshock bubbles\cite{archer15a, omidi10a, turner13a}, are all produced by the interaction between the incident solar wind and suprathermal ($>$100 eV to 100s of keV) ions\cite{burgess12a, burgess13a, wilsoniii16a}.  Short large-amplitude magnetic structures are short duration ($\sim$few to 10s of seconds), nonlinear large amplitude ($\delta B/B$ $>$ 2), monolithic ``magnetic pulsations'' with spatial scales of $\sim$1000 km that can exhibit a soliton-like behavior (i.e., large amplitude fluctuations are fast and spatially narrow)\cite{schwartz92a, mann94a}.  Both hot flow anomalies and foreshock bubbles are localized rarefaction regions surrounded by compression regions that are effectively ``carved out'' by an accumulation of suprathermal ions along a discontinuity in the interplanetary magnetic field.  The difference is that the compression regions for hot flow anomalies are centered on the discontinuity and the discontinuity must interact with the Earth's bow shock, whereas foreshock bubbles form upstream of the discontinuity and the discontinuity need not interact with the Earth's bow shock\cite{omidi07a, turner13a}.  Both hot flow anomalies and foreshock bubbles are several Earth radii in scale (i.e., $>$10,000 km).

%%  Subsection:  Exclusion of Solar Source
{\noindent \textbf{Exclusion of solar source.}}  We first eliminated interplanetary shocks as a possibility by examining the \emph{Wind} shock database at the Harvard Smithsonian for Astrophysics (Online at \textcolor{Blue}{\seqsplit{http://themis.ssl.berkeley.edu/index.shtml}}) finding no interplanetary shocks during any of the energetic electron enhancements.

\begin{figure}[!htb]
  \centering
    {\includegraphics[trim = 0mm 0mm 0mm 0mm, clip, width=170mm]{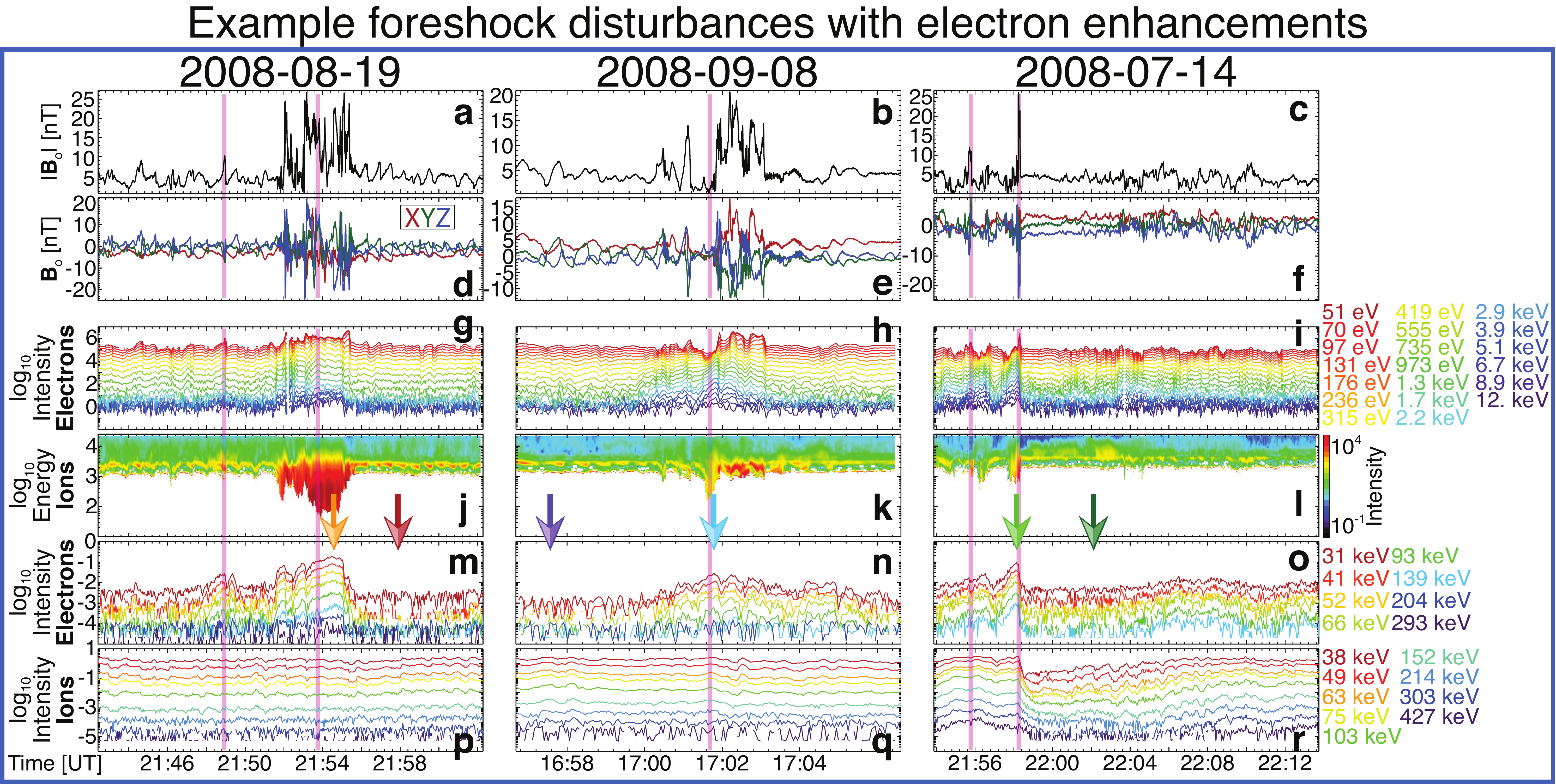}}
    \caption[Example TIFPs with electron enhancements]{\textbf{Three example foreshock disturbances with energetic electron enhancements.}  All data were observed by THEMIS-C, with disturbance centers indicated by the vertical magenta lines.  Magnetic fields are shown in units of nanotesla [nT] in the geocentric solar ecliptic coordinate basis.  Particle data are shown in units of intensity [\# cm$^{-2}$ s$^{-1}$ sr$^{-1}$ eV$^{-1}$] (or flux) as omnidirectional averages in the bulk flow rest frame with uniform color schemes (legends at far right) and vertical axis ranges by row.  The low and high energy electron and high energy ion data are all shown as stacked line plots of intensity versus time, where each line corresponds to a different energy.  The low energy ion data are shown as a dynamic energy spectrogram of energy versus time with a color scale (right of \textbf{l}) for intensity.  \textbf{a}-\textbf{c}, $\lvert \mathbf{B}{\scriptstyle_{o}} \rvert$ [nT] at 4 samples per second.  \textbf{d}-\textbf{f}, $\mathbf{B}{\scriptstyle_{o}}$  [nT] at 4 samples per second.  \textbf{g}-\textbf{i}, low energy electron (i.e., $\sim$50 eV to $\sim$12 keV) intensity.  \textbf{j}-\textbf{l}, low energy ion (i.e., $\sim$10 eV to $\sim$25 keV) intensity.  \textbf{m}-\textbf{o}, high energy electron (i.e., $\sim$30--300 keV) intensity.  \textbf{p}-\textbf{r}, high energy ion (i.e., $\sim$30--430 keV) intensity.  See Methods for details on the calculation of these quantities.}
    \label{fig:exampleTIFPs}
\end{figure}

\indent  We next examine the radio data from the \emph{Wind} and STEREO spacecraft (Extended Data Fig. \ref{fig:WindandSTEREOWAVESRadio}).  There are no clear radio bursts or any evidence of significant radio activity on the sun during any of the four THEMIS foreshock passes.  For comparative purposes, we include a date with clear solar radio bursts (Extended Data Figs \ref{fig:WindandSTEREOWAVESRadio}\textbf{e}, \ref{fig:WindandSTEREOWAVESRadio}\textbf{j}, and \ref{fig:WindandSTEREOWAVESRadio}\textbf{o}).  The enhanced radio intensity near $\sim$200 kHz (Figs \ref{fig:WindandSTEREOWAVESRadio}\textbf{f}--\ref{fig:WindandSTEREOWAVESRadio}\textbf{i}) is most likely auroral kilometric radiation\cite{ergun98b}, which would have no effect on particle observations by THEMIS.  Examination of the solid state telescope particle data from \emph{Wind}\cite{lin95a} (freely available on CDAWeb, see \textbf{Data availability} below) \textcolor{Red}{show} no significant energetic electron enhancements during any of the enhancements observed by THEMIS.  Finally, the electron data \textcolor{Red}{show} no evidence of forward energy dispersion (i.e., higher energies arrive before lower due to a time-of-flight effect) characteristic solar energetic electrons\cite{ergun98e} (Fig. \ref{fig:exampleTIFPs} and Extended Data Figs \ref{fig:THEMISFBPADExample}\textbf{k}--\ref{fig:THEMISFBPADExample}\textbf{n}).  Note that the energetic ions exhibit slightly larger anisotropies than the electrons with anti-earthward intensities generally dominating (see Extended Data Fig. \ref{fig:THEMISFBPADExample}).

%%  Subsection:  Exclusion of Earth's Bow Shock as Source
%%  Discuss different shock acceleration mechanisms and limitations
%%  Discuss inconsistent PADs
%%  Discuss inconsistent aliasing in PADs
%%  Discuss inconsistent energies
%%  Discuss gyroradii vs. FDs scale sizes
%%  etc.
%%
%%    Rule out DSA
{\noindent \textbf{Exclusion of Earth's bow shock as source.}}  The most common shock acceleration mechanisms cited are diffusive shock acceleration\cite{malkov01a}, shock drift acceleration\cite{anagnostopoulos98a, park13a}, and the ``fast Fermi'' mechanism\textcolor{Red}{\cite{wu84b, leroy84a}}.  However, for electron acceleration at the Earth's bow shock we can rule out these mechanisms for the following reasons.  First, though diffusive shock acceleration predicts an isotropic particle distribution, it is more efficient for quasi-parallel ($\theta{\scriptstyle_{Bn}}$ $<$ 45$^{\circ}$) shocks with pre-existing upstream electromagnetic fluctuations and the efficiency increases with particle kinetic energy\cite{caprioli14a, park15a}.  This mechanism also cannot energize electrons below $\sim$100 keV because their Larmor radii are smaller than the gradient scale lengths of the shock \textcolor{Red}{ramp\cite{hobara10a, mazelle10a}} and upstream electromagnetic fluctuations\cite{wilsoniii16a}.  Further, this mechanism predicts an inverse energy dispersion (i.e., lower energies enhance first)\cite{anagnostopoulos86a, sarris87a}, which is not observed in the energetic electron enhancements (Fig. \ref{fig:exampleTIFPs} and Extended Data Fig. \ref{fig:THEMISFBPADExample}).  Thus, diffusive shock acceleration is generally ignored as a mechanism for energizing electrons from thermal to relativistic energies at the Earth's bow shock.

\begin{figure}[!htb]
  \centering
    {\includegraphics[trim = 0mm 0mm 0mm 0mm, clip, width=86mm]{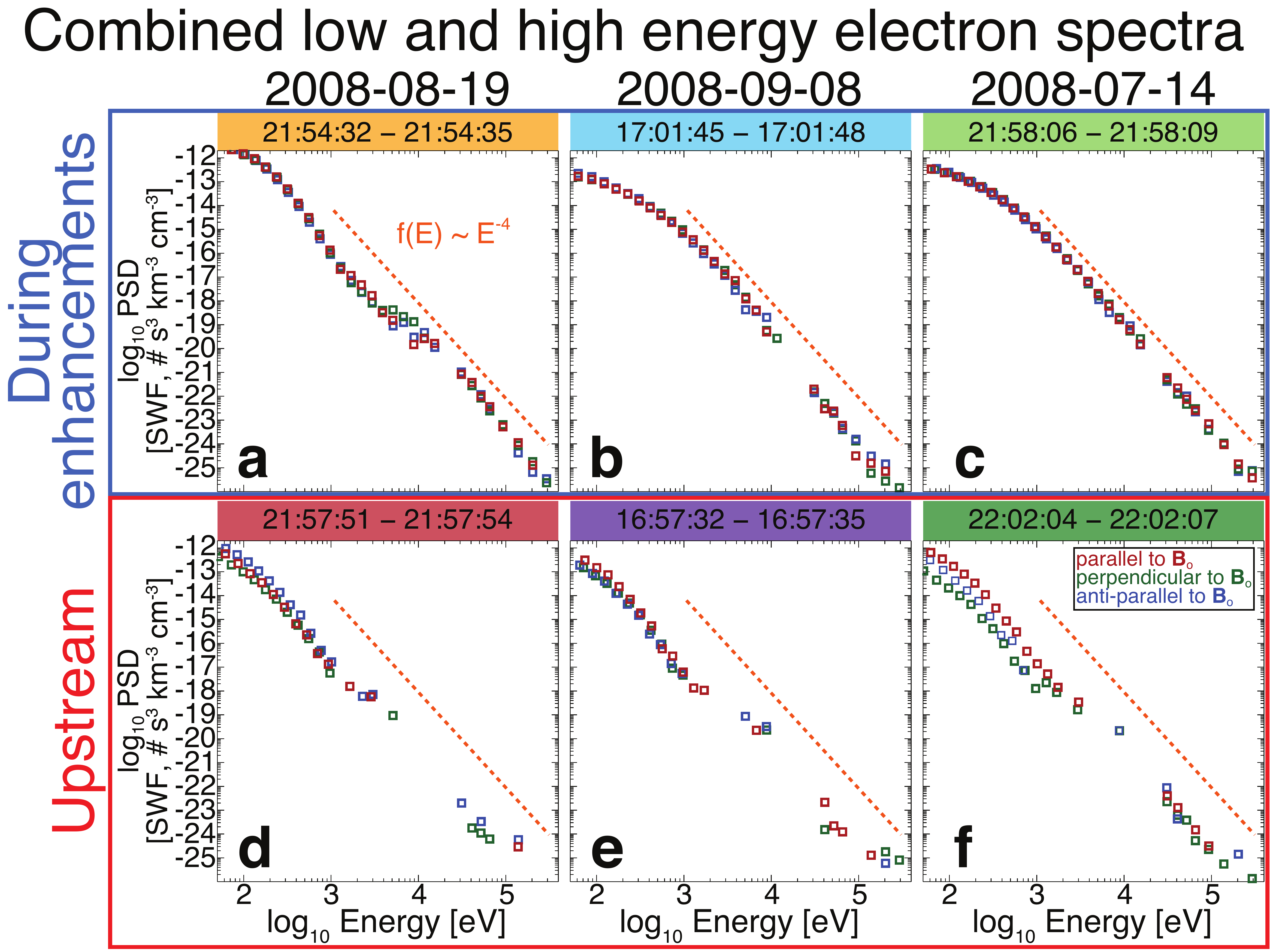}}
    \caption[Example EESA and SSTe in and out of enhancements]{\textbf{Example pitch-angle spectra inside and outside electron enhancement periods for each of the example disturbances in Fig. \ref{fig:exampleTIFPs}.}  One-dimensional directional cuts (with respect to $\mathbf{B}{\scriptstyle_{o}}$) of the merged particle velocity distributions from the low and high energy electron data, in units of phase space density [\# s$^{+3}$ cm$^{-3}$ km$^{-3}$], on log-log plots with uniform horizontal and vertical axis ranges of $\sim$10$^{-26}$--10$^{-12}$ s$^{+3}$ cm$^{-3}$ km$^{-3}$ and $\sim$0.050--400 keV, respectively.  The color-coded boxes above each plot correspond to the color-coded arrows in Fig. \ref{fig:exampleTIFPs}.  Each panel shows cuts parallel (red), perpendicular (green), and anti-parallel (blue) to $\mathbf{B}{\scriptstyle_{o}}$, where missing data points indicate an absence of significant flux.  The orange dashed line shows a power-law (defined in \textbf{a}) for perspective.  \textbf{a}--\textbf{c}, distributions during the peak energetic electron enhancements in Fig. \ref{fig:exampleTIFPs}.  \textbf{d}--\textbf{f}, distributions during periods of inactivity in Fig. \ref{fig:exampleTIFPs}.  See Methods for details on the calculation of these quantities.}
    \label{fig:examplePADs}
\end{figure}

%%    Rule out SDA
\indent  Second, shock drift acceleration predicts anisotropic velocity distributions, perpendicular downstream of shock and field-aligned far upstream of the Earth's bow shock\cite{anagnostopoulos94b, anagnostopoulos09a}, neither of which are observed (Fig. \ref{fig:examplePADs} and Extended Data Fig. \ref{fig:THEMISFBPADExample}).  The mechanism efficiency decreases with increasing ratio of shock speed to $\cos{\theta{\scriptstyle_{Bn}}}$ relative to the particle thermal energy because this increases the minimum energy threshold requirement\cite{guo14b}, where each interaction with the shock can produce energy gains of factors $>$10 for strong quasi-perpendicular shocks\cite{ball01a}.  However, any electron distribution observed far upstream (e.g., near the foreshock disturbances) would be highly anisotropic along the magnetic field streaming away from the shock as previously observed\cite{anderson79a, anderson81a}, inconsistent with the isotropic distributions we observe (Fig. \ref{fig:examplePADs} and Extended Data Fig. \ref{fig:THEMISFBPADExample}).  Thus, we can rule out shock drift acceleration at the Earth's bow shock as a source.

%%    Rule out fast Fermi
\indent  Third, fast Fermi acceleration assumes electrons undergo a single adiabatic reflection -- particle conserves its magnetic moment, $\mu$ $=$ $m{\scriptstyle_{e}} v{\scriptstyle_{\perp}}^{2}/2 B{\scriptstyle_{o}}$ $\sim$ constant, during the reflection, where $m{\scriptstyle_{e}}$ is the electron mass and $v{\scriptstyle_{\perp}}$ is the speed perpendicular to $\mathbf{B}{\scriptstyle_{o}}$ -- and gains energy proportional to the shock speed divided by $\cos{\theta{\scriptstyle_{Bn}}}$\cite{savoini10a, wu84b}.  Previous studies\textcolor{Red}{\cite{anderson79a, anderson81a, leroy84a}} proposed this mechanism as an explanation for the ``thin sheets'' of highly anisotropic (i.e., field-aligned streaming away from Earth's bow shock) energetic electrons.  To satisfy the condition $\mu$ $\sim$ constant, the magnetic gradient scale length must be larger than the particle Larmor radius.  Further, for significant electron energy gains this mechanism requires either very large shock speeds compared to typical electron thermal speeds (i.e., $\sim$1500--3000 km/s) or $\theta{\scriptstyle_{Bn}}$ $\gtrsim$ 88$^{\circ}$.  Since the Earth's bow shock is very slow (i.e., typically $\sim$100--500 km/s), the Larmor radii of electrons $\gtrsim$ few hundred eV are comparable to the shock ramp \textcolor{Red}{thickness\cite{hobara10a, mazelle10a}}, and this mechanism can only energize electrons with large pitch-angles, fast Fermi acceleration is not expected to produce energies beyond several 10s of keV\cite{gosling89b, sarris85a, savoini10a, wu84b}.  We thus rule out the Earth's bow shock as the source of these energetic electron enhancements.

%%  Subsection:  Exclusion of Earth's Magnetosphere as Source
%%    Rule out due to PADs
{\noindent \textbf{Exclusion of Earth's magnetosphere as source.}}  As discussed in the main article, the observed energetic electron distributions are not highly anisotropic along the magnetic field streaming away from the Earth as previously reported in studies arguing for a magnetospheric source\cite{anagnostopoulos94b, krimigis78a, kronberg11a, sarris76a, sarris87a}.  These studies observed enhanced geomagnetic activity in association with bursts of energetic electrons, where they argued that substorms -- a fundamental mode of the terrestrial magnetosphere resulting in magnetospheric circulation/flows and enhanced auroral activity\cite{angelopoulos08a} -- led to an increased rate of magnetospheric ``leakage.''  During substorms geostationary spacecraft can observe intense and rapid changes in 10s of keV to MeV electron fluxes\cite{angelopoulos08a}.  One measure of substorm activity can be given by the well known AE indices\cite{mann08a}.  These previous studies defined enhanced geomagnetic activity as an AE index $>$ 200 nT.

%%  Discuss inconsistent AE
\indent  Extended Data Fig. \ref{fig:THEMISAEIndices} shows the AE indices with color-coded bars indicating the time ranges for the example foreshock disturbances in Fig. \ref{fig:exampleTIFPs}.  The AE index (Extended Data Figs \ref{fig:THEMISAEIndices}\textbf{q}--\ref{fig:THEMISAEIndices}\textbf{t}) was $<$ 200 nT for $>$1 hour prior to and during the three example disturbances with energetic electron enhancements (Figs \ref{fig:exampleTIFPs}\textbf{m}--\ref{fig:exampleTIFPs}\textbf{o}).  Thus, we can rule out a magnetospheric source due to the inconsistent pitch-angle distributions and lack of enhanced geomagnetic activity.

%%  Discuss Paschmann et al. [1988]
\indent  \textcolor{Red}{One previous study\cite{paschmann88a} did find nearly isotropic energetic electrons ($\sim$70--200 keV) within hot flow anomalies.  However, the authors explicitly state these are not a suprathermal tail of the thermal electrons and that they originated from the magnetosphere.  There are several important differences with our results:  (1) we observe a single power-law from 100s of eV to $\geq$140 keV in many enhancements, suggesting a common acceleration mechanism;  (2) we always observe ions above $\sim$10 keV with every foreshock disturbance;  (3) we observe most electron enhancements (8/10) within short large-amplitude magnetic structures and foreshock bubbles, both of which are disconnected from the bow shock; and (4) the ``magnetic bottle'' model proposed to explain isotropy in the previous study\cite{paschmann88a} would not work for short large-amplitude magnetic structures because they contract as they evolve, not expand.}

%%  Subsection:  Instrument Details
{\noindent \textbf{Instrument details.}}  Quasi-static (i.e., finite gain from zero up to Nyquist frequency) vector magnetic field measurements ($\mathbf{B}{\scriptstyle_{o}}$) were obtained using the fluxgate magnetometer\cite{auster08a} at 4 and 128 samples per second.  The data are presented in units of nanotesla [nT] in the geocentric solar ecliptic coordinate basis.

\indent  Particle data are stored as velocity distribution functions covering $4 \ \pi$ steradian over an energy range defined by instrument design.  The electrostatic analyzers\cite{mcfadden08a, mcfadden08b}, onboard each THEMIS\cite{angelopoulos08a} spacecraft, detect particles using anodes placed behind microchannel plate detectors and cover an energy range of few eV to over 25 keV.  The number and placement of the anodes determines the poloidal/latitudinal angular resolution, which is usually $\Delta \theta$ $\sim$ 22.5$^{\circ}$ for both the electron and ion electrostatic analyzers (Note that $\Delta \theta$ can be as low as $\sim$5$^{\circ}$ in some ion instrument modes.).  The azimuthal/longitudinal resolution, $\Delta \phi$, is limited by the spacecraft spin rate and instrument design and mode of operation, but is generally $\sim$11.25$^{\circ}$.  The energy resolution, $\Delta E/E$, is defined by the instrument design and mode of operation but is generally $\sim$20\% for both electrostatic analyzers.

\indent  At higher energies (i.e., $\sim$30 keV to over 500 keV), data from the electron and ion solid state telescopes\cite{ni11b, turner13b} onboard each THEMIS spacecraft were used.  Each detector is comprised of two identical telescopes mounted at different angles on the side of the spacecraft body\cite{angelopoulos08a}.  The angular and energy resolution is usually $\Delta \theta$ $\sim$ 30$^{\circ}$--40$^{\circ}$, $\Delta \phi$ $\sim$ 22.5$^{\circ}$, and $\Delta E/E$ $\sim$ $\sim$ 30\%.  For instance, the $\sim$293 keV energy bin actually includes energetic electrons with $\sim$249--337 keV energies.

\indent  All particle data presented herein, except for the high energy ions, \textcolor{Red}{were} taken \textcolor{Red}{while in} burst \textcolor{Red}{mode,} which has a time resolution equal to the spacecraft spin period (i.e., $\sim$3 seconds).  Even though the high energy ion data were measured in a different mode, the time resolution is still the spin period for the intervals presented.

%%  Subsection:  Unit Conversion
{\noindent \textbf{Unit conversion.}}  The raw data are measured in units of counts, which correspond to the number of events with a pulse height exceeding a defined threshold specific to each instrument.  Conversion to intensity and/or phase space density requires knowledge of the instruments efficiency\cite{bordoni71a, goruganthu84a}, deadtime\cite{meeks08a, schecker92a}, accumulation time, and optical geometric factor\cite{curtis89a, paschmann98a, wuest07b}.

\indent  Particle intensity is defined as:  number of particles, per unit area, per unit solid angle, per unit time, per unit energy (e.g., \# cm$^{-2}$ s$^{-1}$ sr$^{-1}$ eV$^{-1}$).  This unit is not a Lorentz invariant, thus it requires one taking into account the Compton-Getting effect.  Phase space density is defined as:  number of particles, per unit spatial volume, per unit ``velocity volume'' (e.g., \# s$^{+3}$ cm$^{-3}$ km$^{-3}$).  This unit is a Lorentz invariant under conditions when phase space is incompressible (i.e., when Liouville's theorem reduces to $df/dt$ $=$ 0), which is true for most cases in in situ space observations\cite{paschmann98a}.

\indent  The exact details of the unit conversion can be found in the THEMIS calibration software, called SPEDAS, found at: \textcolor{Blue}{\seqsplit{http://themis.ssl.berkeley.edu/index.shtml}}.

%%  Subsection:  Reference Frames and Coordinate Systems
{\noindent \textbf{Reference frames and coordinate systems.}}  All particle data shown herein has been transformed from the spacecraft to the bulk flow reference frame using a relativistically correct Lorentz transformation.  The distributions were converted into units of phase space density prior to any frame transformation, thus any anisotropies due to the Compton-Getting effect\cite{compton35a, ipavich74a} have been removed.

\noindent  Reference frame transformations were performed through the following steps:
\begin{enumerate}[itemsep=0pt,parsep=0pt,topsep=0pt]
  \item  bulk flow velocities were determined from the first velocity moment of the low energy ($<$30 keV) ion velocity distributions\cite{wilsoniii14a};
  \item  all particle distributions were converted from raw counts to phase space density;
  \item  particle distributions were then transformed into the bulk flow rest frame using a standard Lorentz velocity transformation;
  \item  the particle distributions shown in units of intensity were converted to phase space density prior to the transformation and then back to intensity.
\end{enumerate}

\indent  Direction-dependent spectra, as opposed to omnidirectional averages (i.e., average over all solid angles), called pitch-angle distributions were calculated through the following steps:
\begin{enumerate}[itemsep=0pt,parsep=0pt,topsep=0pt]
  \item  particle distributions were transformed into the bulk flow rest frame as described above;
  \item  construct particle velocity unit vector for the $m^{th}$ particle distribution, $\hat{\mathbf{v}}{\scriptstyle_{i,j,k}}^{m}$, for the $k^{th}$ energy bin from the $i^{th}$ latitude and $j^{th}$ longitude detector look directions described in \textbf{Instrument details} above;
  \item  define the projection angle, $\alpha{\scriptstyle_{i,j,k}}^{m}$, between $\hat{\mathbf{v}}{\scriptstyle_{i,j,k}}^{m}$ and the respective orientation unit vector, $\hat{\mathbf{u}}^{m}$ (e.g., $\mathbf{B}{\scriptstyle_{o}}$/$\lvert \mathbf{B}{\scriptstyle_{o}} \rvert$), at the measurement time of the $m^{th}$ particle distribution, where $\alpha{\scriptstyle_{i,j,k}}^{m}$ $=$ $\cos^{-1}{\left( \hat{\mathbf{v}}{\scriptstyle_{i,j,k}}^{m} \cdot \hat{\mathbf{u}}^{m} \right)}$; and
  \item  define three cuts for the $m^{th}$ particle distribution by averaging data within $\pm$22.5$^{\circ}$ of the parallel, perpendicular, and anti-parallel directions defined by $\hat{\mathbf{u}}^{m}$.
\end{enumerate}

\noindent  The two relevant directions about which we oriented the particle distributions are the local quasi-static magnetic field vector, $\mathbf{B}{\scriptstyle_{o}}$, and the spacecraft-to-Earth unit vector, $\hat{\mathbf{e}}{\scriptstyle_{SC}}$, at the time of each distribution.  Note that the formal definition of a pitch-angle distribution is constructed only with respect to $\mathbf{B}{\scriptstyle_{o}}$ but we use the term here for both orientations for brevity.

\indent  The high time resolution equivalent of the above algorithm involves only a few differences described below.
\begin{enumerate}[itemsep=0pt,parsep=0pt,topsep=0pt]
  \item  Instead of using a single $\hat{\mathbf{u}}^{m}$ for the $m^{th}$ particle distribution, we define $\hat{\mathbf{u}}{\scriptstyle_{i,j,k}}^{m}$ for the $k^{th}$ energy bin from the $i^{th}$ latitude and $j^{th}$ longitude detector look directions of the $m^{th}$ particle distribution.
  \item  Now the pitch-angles are defined as $\alpha{\scriptstyle_{i,j,k}}^{m}$ $=$ $\cos^{-1}{\left( \hat{\mathbf{v}}{\scriptstyle_{i,j,k}}^{m} \cdot \hat{\mathbf{u}}{\scriptstyle_{i,j,k}}^{m} \right)}$.
\end{enumerate}

\noindent  The result is a pitch-angle distribution with fewer aliasing effects due to the use of a single $\hat{\mathbf{u}}^{m}$ averaged over the duration of the $m^{th}$ particle distribution (Extended Data Fig. \ref{fig:exampleCompareLTHTBo}).

\indent  The exact details of the rotations and frame transformations can be found in the additional analysis software at:  \textcolor{Blue}{\seqsplit{https://github.com/lynnbwilsoniii/{wind\_3dp\_pros}}}.

{\noindent \textbf{Particle data presentation.}}  All particle distributions are presented in the bulk flow rest frame in a physically significant coordinate basis, e.g., magnetic field-aligned coordinates.  We define the bulk flow reference frame as described above (in \textbf{Reference frames and coordinate systems}).  The energy ranges listed above (in \textbf{Instrument details}) are the measured midpoint kinetic energies in the spacecraft frame of reference.  Energy values will change under any Lorentz transformation.  Thus, the energies for the ion electrostatic analyzer data are not constant in time and show a large variability owing to the large variability of the bulk flow near foreshock disturbances.  In contrast, the low energy electron data above $\sim$50 eV and all solid state telescope data suffer little energy change under these Lorentz transformations, thus can be approximated as constant in time.  Therefore, we show the low and high energy electron and high energy ion data as stacked line plots vs. time where each line corresponds to a different energy and the low energy ion data are presented as a dynamic energy spectrogram of energy vs. time with the color scale indicating the particle intensity.  Note that while the electron solid state telescope can measure electrons to $>$400 keV, for only the 2008-08-19 event in Fig. \ref{fig:exampleTIFPs} were significant fluxes observed in energy bins $>$140 keV.

{\noindent \textbf{Data availability.}}  The THEMIS data used in this paper is publicly available at: \\
\textcolor{Blue}{\seqsplit{http://themis.ssl.berkeley.edu/index.shtml}}.

\noindent  The \emph{Wind} and STEREO radio data were taken from the S/WAVES website at: \\ 
\textcolor{Blue}{\seqsplit{http://swaves.gsfc.nasa.gov/data\_access.html}}.

\noindent  Solar wind data was taken from the OMNI data products found on CDAWeb at: \\
\textcolor{Blue}{\seqsplit{http://cdaweb.gsfc.nasa.gov}}.

\noindent  The \emph{Wind} interplanetary shock list can be found at: \\
\textcolor{Blue}{\seqsplit{https://www.cfa.harvard.edu/shocks/wi\_data/}}.

{\noindent \textbf{Code availability.}}  The THEMIS instrument calibration software, called SPEDAS, can be found at: \\ 
\textcolor{Blue}{\seqsplit{http://themis.ssl.berkeley.edu/index.shtml}};

\noindent and additional analysis software can be found at: \\
\textcolor{Blue}{\seqsplit{https://github.com/lynnbwilsoniii/{wind\_3dp\_pros}}}.

%%\clearpage
%%----------------------------------------------------------------------------------------
%%  Bibliography
%%----------------------------------------------------------------------------------------

\clearpage
\onecolumn
%%----------------------------------------------------------------------------------------
%%  Figures
%%----------------------------------------------------------------------------------------

%%++++++++++++++++++++++++++++++++++++++++++++++++++++++++++++++++++++++++++++++++++++++++
%% Image:  Example TIFPs with electron enhancements
%%++++++++++++++++++++++++++++++++++++++++++++++++++++++++++++++++++++++++++++++++++++++++
%%++++++++++++++++++++++++++++++++++++++++++++++++++++++++++++++++++++++++++++++++++++++++
%% Image:  Example TIFPs with electron enhancements
%%++++++++++++++++++++++++++++++++++++++++++++++++++++++++++++++++++++++++++++++++++++++++

%%++++++++++++++++++++++++++++++++++++++++++++++++++++++++++++++++++++++++++++++++++++++++
%% Image:  Example EESA and SSTe in and out of enhancements
%%++++++++++++++++++++++++++++++++++++++++++++++++++++++++++++++++++++++++++++++++++++++++
%%++++++++++++++++++++++++++++++++++++++++++++++++++++++++++++++++++++++++++++++++++++++++
%% Image:  Example EESA and SSTe in and out of enhancements
%%++++++++++++++++++++++++++++++++++++++++++++++++++++++++++++++++++++++++++++++++++++++++

\clearpage
%%----------------------------------------------------------------------------------------
%%  Extended Data Figures
%%----------------------------------------------------------------------------------------
\setcounter{figure}{0}  %%  Start counter back at zero
\renewcommand{\figurename}{Extended Data Figure}  %%  Change figure name label

%%++++++++++++++++++++++++++++++++++++++++++++++++++++++++++++++++++++++++++++++++++++++++
%% Image:  THEMIS Foreshock Orbits
%%++++++++++++++++++++++++++++++++++++++++++++++++++++++++++++++++++++++++++++++++++++++++
\begin{figure}[!htb]
  \centering
    {\includegraphics[trim = 0mm 0mm 0mm 0mm, clip, width=170mm]{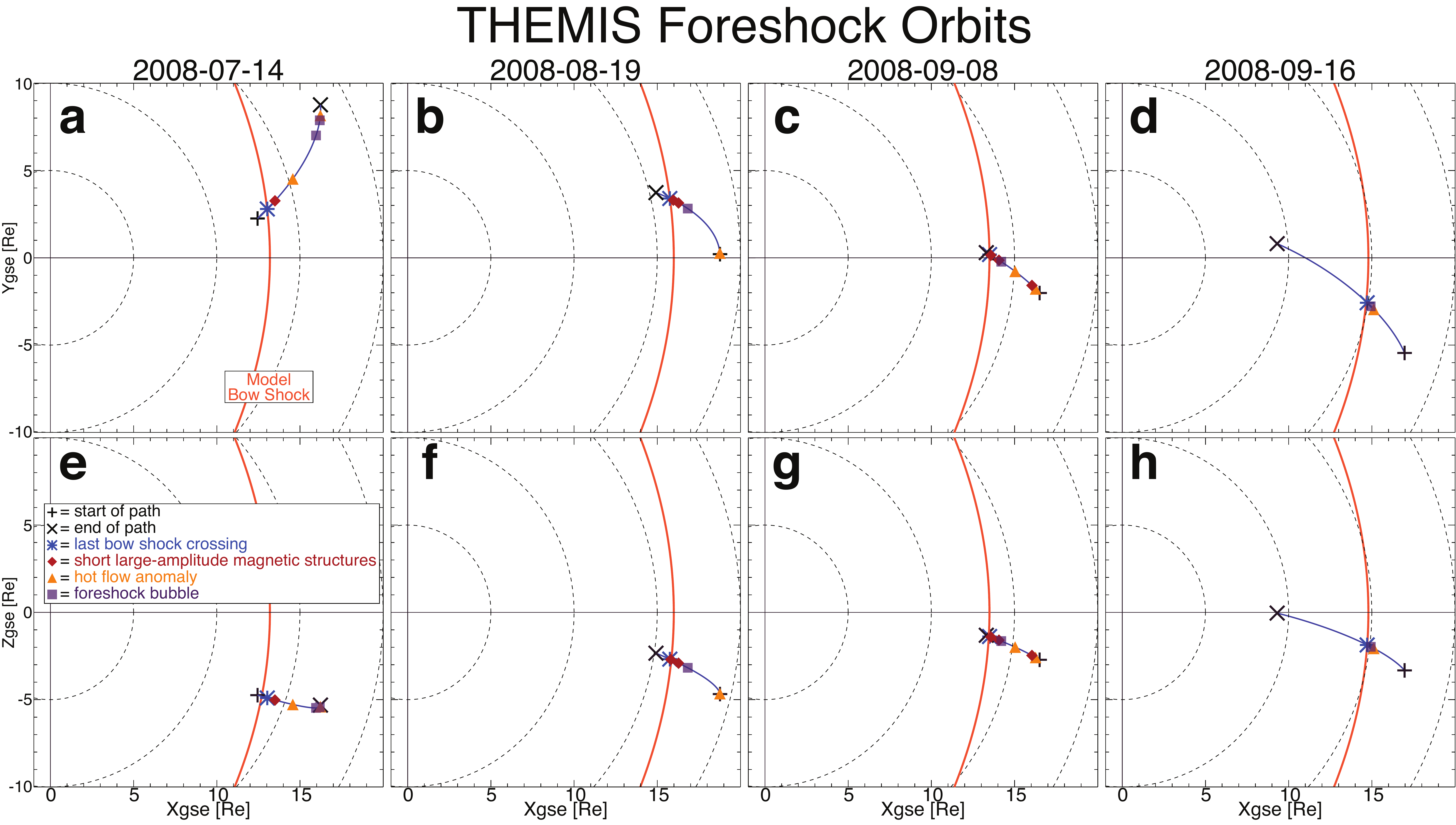}}
    \caption[THEMIS Foreshock Orbits]{\textbf{THEMIS-C spacecraft foreshock orbits.}  Orbital passes of the THEMIS-C spacecraft for the four dates (columns) examined in this study.  \textbf{a}--\textbf{d}, spacecraft orbital path in the Y vs. X geocentric solar ecliptic plane.  \textbf{e}--\textbf{h}, spacecraft orbital path in the Z vs. X geocentric solar ecliptic plane.  The locations of the foreshock disturbances, start/end of pass, and last bow shock crossing are marked by unique symbols along path (legend in \textcolor{Red}{\textbf{e}}).  The solid orange line shows the empirical model bow shock\cite{slavin81a}.}
    \label{fig:THEMISOrbits}
\end{figure}
%%++++++++++++++++++++++++++++++++++++++++++++++++++++++++++++++++++++++++++++++++++++++++
%% Image:  THEMIS Foreshock Orbits
%%++++++++++++++++++++++++++++++++++++++++++++++++++++++++++++++++++++++++++++++++++++++++

%%++++++++++++++++++++++++++++++++++++++++++++++++++++++++++++++++++++++++++++++++++++++++
%% Image:  Example TIFPs without electron enhancements
%%++++++++++++++++++++++++++++++++++++++++++++++++++++++++++++++++++++++++++++++++++++++++
\begin{figure}[!htb]
  \centering
    {\includegraphics[trim = 0mm 0mm 0mm 0mm, clip, width=170mm]{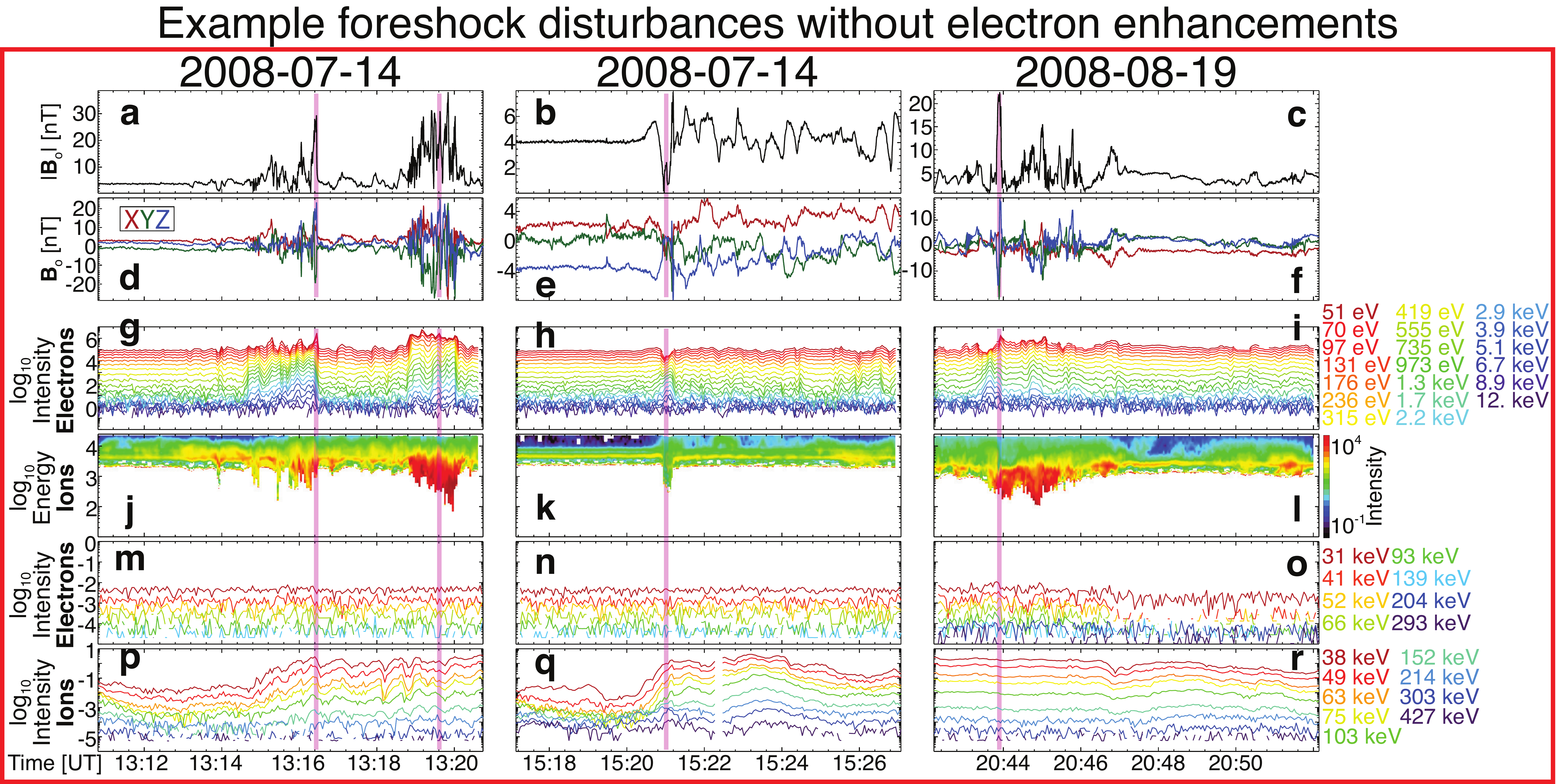}}
    \caption[Example TIFPs without electron enhancements]{\textbf{Three example foreshock disturbances without energetic electron enhancements.}  This figure has the same format as Fig. \ref{fig:exampleTIFPs} \textcolor{Red}{with magnetic fields shown in nT and particle data in intensity  [\# cm$^{-2}$ s$^{-1}$ sr$^{-1}$ eV$^{-1}$]}.  \textbf{a}-\textbf{c}, $\lvert \mathbf{B}{\scriptstyle_{o}} \rvert$ [nT] at 4 samples per second.  \textbf{d}-\textbf{f}, $\mathbf{B}{\scriptstyle_{o}}$  [nT] at 4 samples per second.  \textbf{g}-\textbf{i}, low energy electron (i.e., $\sim$50 eV to $\sim$12 keV) intensity.  \textbf{j}-\textbf{l}, low energy ion (i.e., $\sim$10 eV to $\sim$25 keV) intensity.  \textbf{m}-\textbf{o}, high energy electron (i.e., $\sim$30--300 keV) intensity.  \textbf{p}-\textbf{r}, high energy ion (i.e., $\sim$30--430 keV) intensity.  See Methods for details on the calculation of these quantities.}
    \label{fig:exampleTIFPswithoutee}
\end{figure}
%%++++++++++++++++++++++++++++++++++++++++++++++++++++++++++++++++++++++++++++++++++++++++
%% Image:  Example TIFPs without electron enhancements
%%++++++++++++++++++++++++++++++++++++++++++++++++++++++++++++++++++++++++++++++++++++++++

%%++++++++++++++++++++++++++++++++++++++++++++++++++++++++++++++++++++++++++++++++++++++++
%% Image:  Example Compare Low and High Time Resolution Bo
%%++++++++++++++++++++++++++++++++++++++++++++++++++++++++++++++++++++++++++++++++++++++++
\begin{figure}[!htb]
  \centering
    {\includegraphics[trim = 0mm 0mm 0mm 0mm, clip, width=150mm]{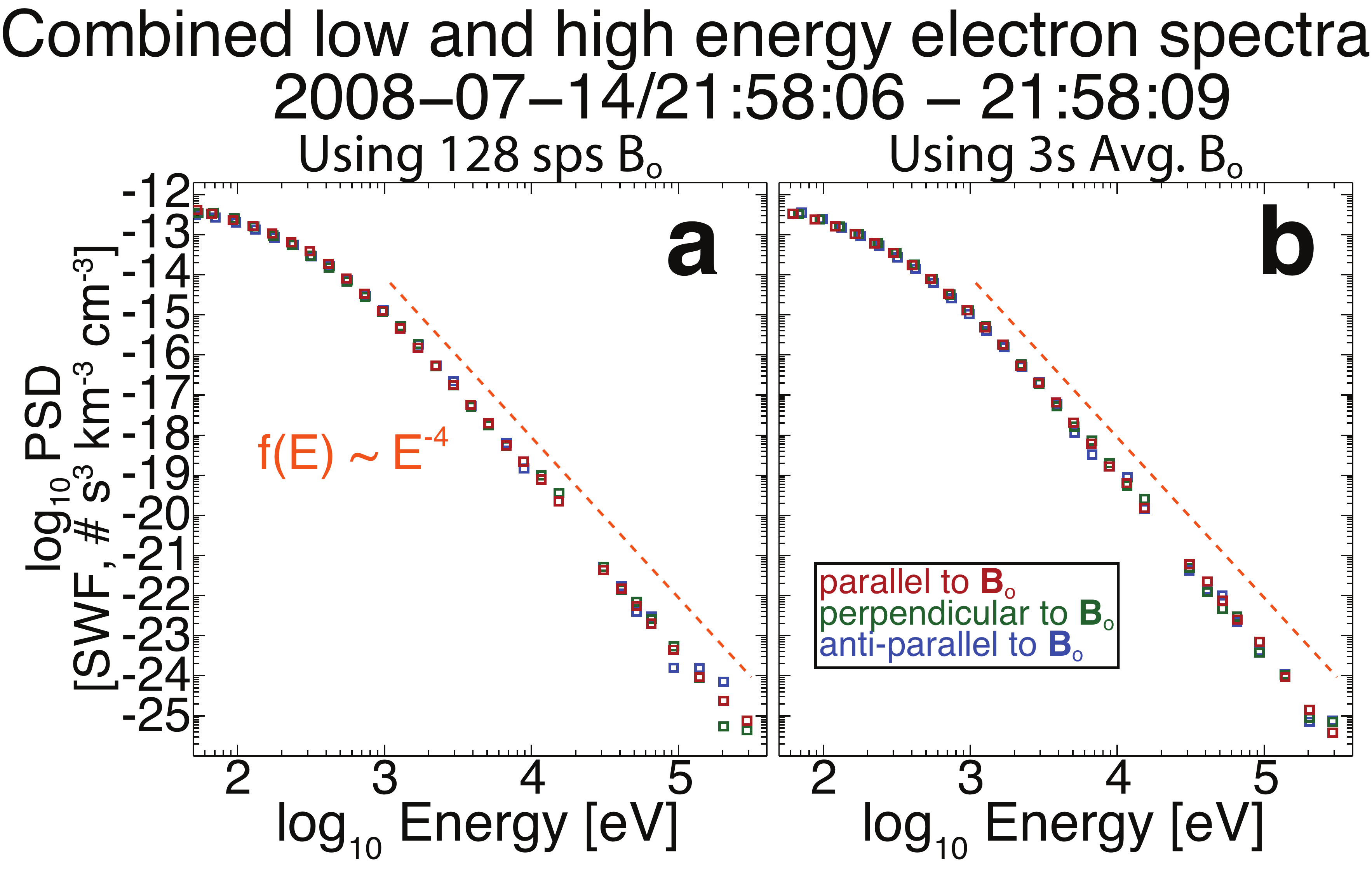}}
    \caption[Example Compare Low and High Time Resolution Bo]{\textbf{Comparison of a single particle distribution constructed from low and high time resolution fields.}  The pitch-angle distribution corresponds to that in Fig. \ref{fig:examplePADs}\textbf{c} with the same format.  \textbf{a}, pitch-angles constructed using $\sim$128 samples per second $\mathbf{B}{\scriptstyle_{o}}$.  \textbf{b}, pitch-angles constructed using \textcolor{Red}{$\sim$3 second averaged} $\mathbf{B}{\scriptstyle_{o}}$.  See Methods for details on the calculation of these quantities.}
    \label{fig:exampleCompareLTHTBo}
\end{figure}
%%++++++++++++++++++++++++++++++++++++++++++++++++++++++++++++++++++++++++++++++++++++++++
%% Image:  Example Compare Low and High Time Resolution Bo
%%++++++++++++++++++++++++++++++++++++++++++++++++++++++++++++++++++++++++++++++++++++++++

%%++++++++++++++++++++++++++++++++++++++++++++++++++++++++++++++++++++++++++++++++++++++++
%% Image:  Wind and STEREO Radio Spectra
%%++++++++++++++++++++++++++++++++++++++++++++++++++++++++++++++++++++++++++++++++++++++++
\begin{figure}[!htb]
  \centering
    {\includegraphics[trim = 0mm 0mm 0mm 0mm, clip, width=170mm]{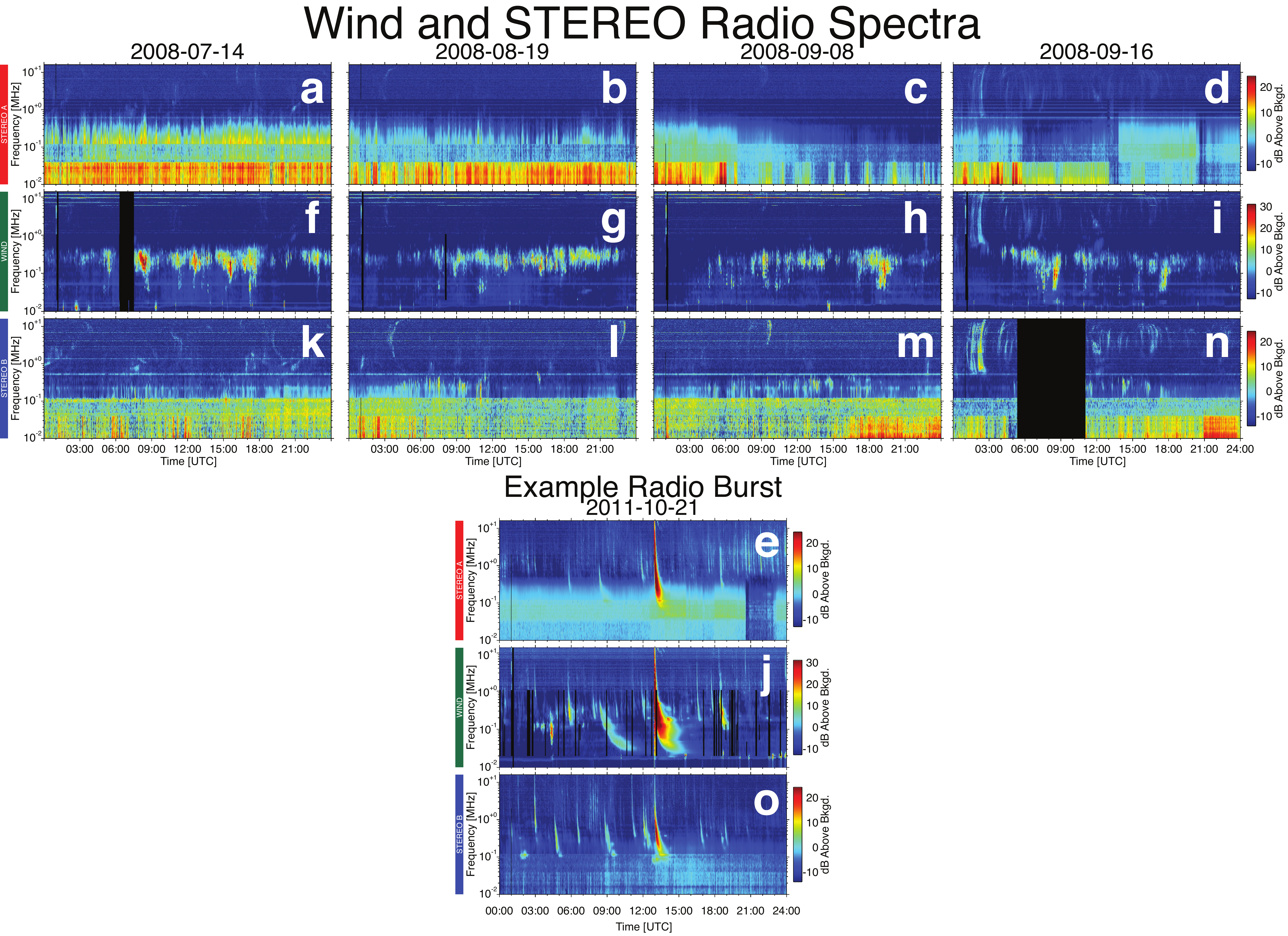}}
    \caption[\emph{Wind} and STEREO Radio Spectra]{\textbf{\emph{Wind} and STEREO spacecraft daily radio spectra plots.}  Radio wave spectra, from the \emph{Wind}\cite{bougeret95a} (panels \textbf{f}--\textbf{j}) and STEREO\cite{bougeret08a} Ahead (panels \textbf{a}--\textbf{e}) and Behind (panels \textbf{k}--\textbf{o}) spacecraft, shown as frequency vs. time plots of daily time- and frequency-averaged spectral intensity plots of decibels above background.  The frequencies shown range from $\sim$10 kHz to $\sim$10 MHz.  \textbf{e}, \textbf{j}, \textbf{o}, show an example radio burst occurring shortly after 12:00 UT on 2011-10-21 seen by all three spacecraft.  Comparison with the example radio burst clearly shows that there is insigificant solar activity for the four foreshock crossings examined in this study.}
    \label{fig:WindandSTEREOWAVESRadio}
\end{figure}
%%++++++++++++++++++++++++++++++++++++++++++++++++++++++++++++++++++++++++++++++++++++++++
%% Image:  Wind and ACE Radio Spectra
%%++++++++++++++++++++++++++++++++++++++++++++++++++++++++++++++++++++++++++++++++++++++++

%%++++++++++++++++++++++++++++++++++++++++++++++++++++++++++++++++++++++++++++++++++++++++
%% Image:  THEMIS Example PADs
%%++++++++++++++++++++++++++++++++++++++++++++++++++++++++++++++++++++++++++++++++++++++++
\begin{figure}[!htb]
  \centering
    {\includegraphics[trim = 0mm 0mm 0mm 0mm, clip, width=170mm]{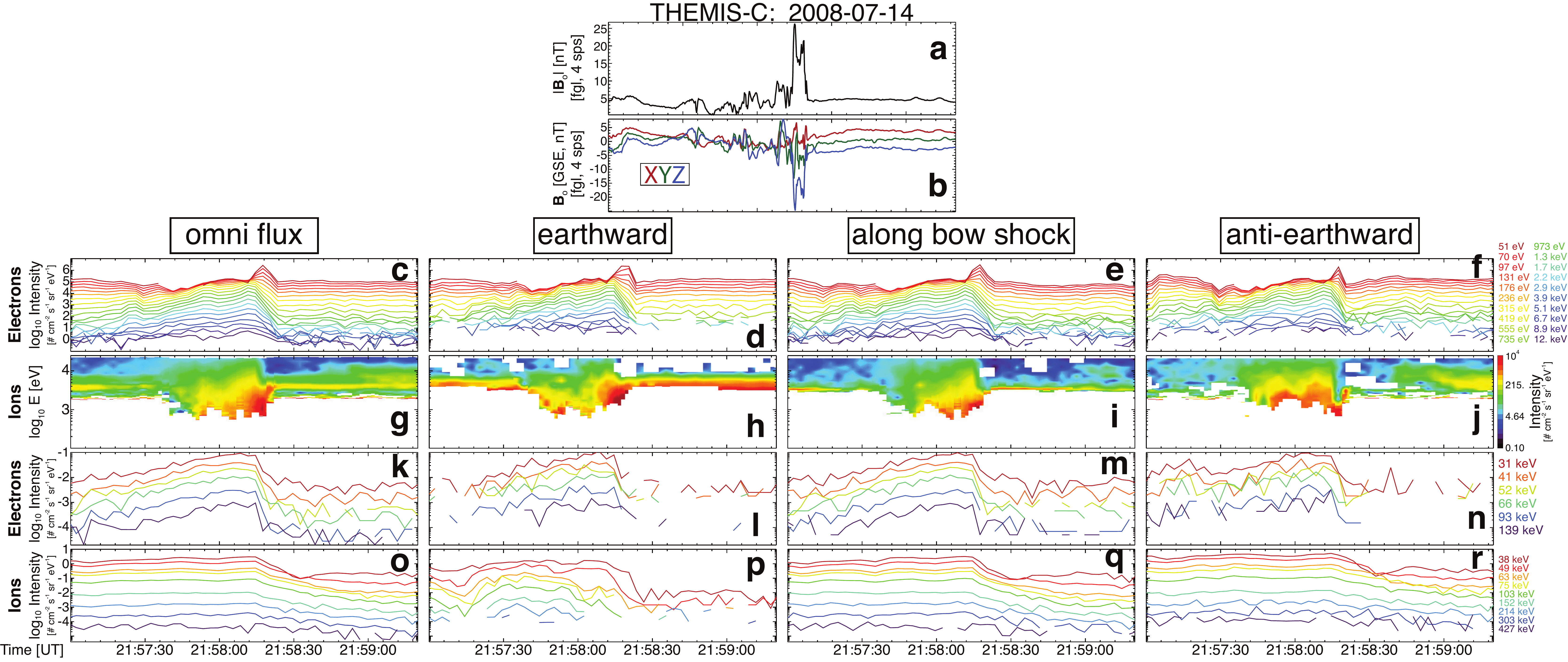}}
    \caption[THEMIS Example PADs]{\textbf{THEMIS-C example electron enhancement observed with the foreshock bubble in \textcolor{Red}{Fig. \ref{fig:exampleTIFPs}} split by direction.}  All particle data in this figure are shown in units of intensity [\# cm$^{-2}$ s$^{-1}$ sr$^{-1}$ eV$^{-1}$] (or flux or number flux) at spin period resolution ($\sim$3 seconds) and magnetic fields are shown in units of nanoteslas [nT] at 4 samples per second.  The vector about which the particle velocity distributions are oriented is spacecraft-to-Earth unit vector, $\hat{\mathbf{e}}{\scriptstyle_{SC}}$.  \textcolor{Red}{The tick marks for \textbf{a} and \textbf{b} are the same as for the rest of the panels.}  \textbf{a} and \textbf{b}, $\lvert \mathbf{B}{\scriptstyle_{o}} \rvert$ and vector components of $\mathbf{B}{\scriptstyle_{o}}$ [nT].  \textbf{c}--\textbf{f}, low energy electron (i.e., $\sim$50 eV to $\sim$12 keV) intensity.  \textbf{g}--\textbf{j}, low energy ion (i.e., $\sim$10 eV to $\sim$25 keV) intensity.  \textbf{k}--\textbf{n}, high energy electron (i.e., $\sim$30--140 keV) intensity.  \textbf{o}--\textbf{r}, high energy ion (i.e., $\sim$30--430 keV) intensity.}
    \label{fig:THEMISFBPADExample}
\end{figure}
%%++++++++++++++++++++++++++++++++++++++++++++++++++++++++++++++++++++++++++++++++++++++++
%% Image:  THEMIS Example PADs
%%++++++++++++++++++++++++++++++++++++++++++++++++++++++++++++++++++++++++++++++++++++++++

%%++++++++++++++++++++++++++++++++++++++++++++++++++++++++++++++++++++++++++++++++++++++++
%% Image:  THEMIS Foreshocks with AE Indices
%%++++++++++++++++++++++++++++++++++++++++++++++++++++++++++++++++++++++++++++++++++++++++
\begin{figure}[!htb]
  \centering
    {\includegraphics[trim = 0mm 0mm 0mm 0mm, clip, width=170mm]{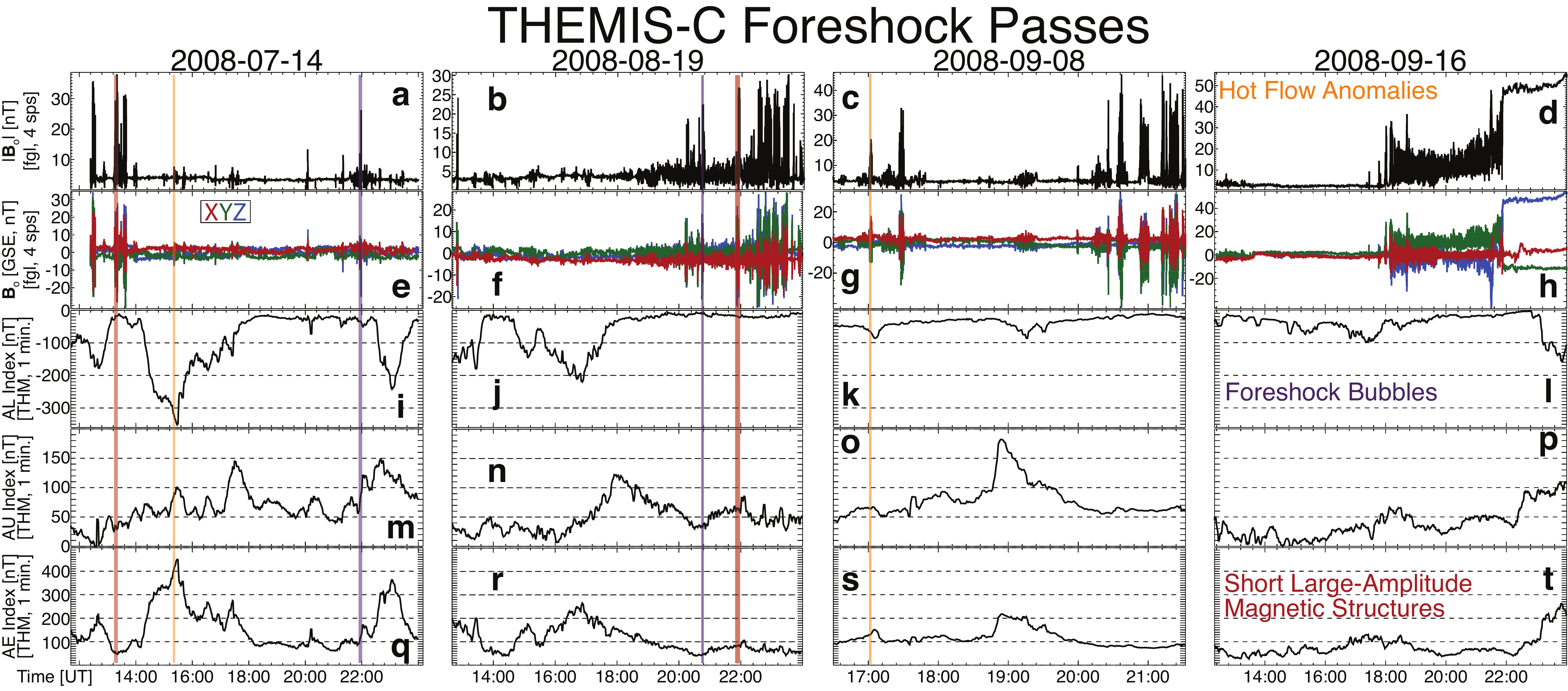}}
    \caption[THEMIS Foreshocks with AE Indices]{\textbf{THEMIS-C foreshock overviews with AL, AU, and AE indices.}  All magnetic field data in this figure are shown in units of nanoteslas [nT], with spacecraft fields measured at 4 samples per second by the fluxgate magnetometer\cite{auster08a} and ground data at 60 second resolution by the THEMIS ground magnetometer arrays\cite{mann08a}.  \textbf{a}--\textbf{d}, $\lvert \mathbf{B}{\scriptstyle_{o}} \rvert$ [nT] at 4 samples per second.  \textbf{e}--\textbf{h}, $\mathbf{B}{\scriptstyle_{o}}$ [nT] at 4 samples per second.  \textbf{i}--\textbf{l}, AL index [nT].  \textbf{m}--\textbf{p}, AU index [nT].  \textbf{q}--\textbf{t}, AE index [nT].  The vertical shaded regions (labels in \textbf{d}, \textbf{l}, and \textbf{t}) show intervals corresponding to the \textcolor{Red}{example} foreshock disturbances shown in Fig. \ref{fig:exampleTIFPs} \textcolor{Red}{and Extended Data Fig. \ref{fig:exampleTIFPswithoutee}}.}
    \label{fig:THEMISAEIndices}
\end{figure}
%%++++++++++++++++++++++++++++++++++++++++++++++++++++++++++++++++++++++++++++++++++++++++
%% Image:  THEMIS Foreshocks with AE Indices
%%++++++++++++++++++++++++++++++++++++++++++++++++++++++++++++++++++++++++++++++++++++++++

\end{document}